\documentclass[fleqn,twoside]{article}
\usepackage{espcrc2}

\usepackage{graphicx}
\usepackage[figuresright]{rotating}

\newcommand{\AmS}{{\protect\the\textfont2
  A\kern-.1667em\lower.5ex\hbox{M}\kern-.125emS}}

\usepackage{epsfig}
\input{pstricks.tex}

\def\bi{\begin{itemize}}
\def\ei{\end{itemize}}
\let\ib=\item

\def\etc{{\it etc.}}
\def\eg{{\it e.g.}}
\def\vs{{\it vs.}}
\def\etal{{\it et al.}}
\def\ie{{\it i.e.}}

\def\BR{{\cal B}}

\def\nubar{\overline{\nu}}

\def\antibar#1{{\overline{#1}}}

\def\nutau{{\nu_\tau}}

\def\nutaubar{\antibar{\nu}_\tau}
\newcommand{\pizero}{\mbox{$\pi^{0}\,$}}
\def\hnt{h_{\nu_\tau}}

\title{ CLEO Contributions to Tau Physics}

\author{Alan. J.~Weinstein
        \address{California Institute of Technology, \\
        Pasadena, CA 91125, USA \\
        Representing the CLEO Collaboration}
        \thanks{Work supported by the US Department of Energy 
                and National Science Foundation.}
       }
       
\begin{document}

\begin{abstract}
We review many of the contributions of the CLEO experiment 
to tau physics. Topics discussed are:
branching fractions for major decay modes
and tests of lepton universality;
rare decays; forbidden decays;
Michel parameters and spin physics;
hadronic sub-structure and resonance parameters;
the tau mass, tau lifetime, and tau neutrino mass;
searches for CP violation in tau decay;
tau pair production, dipole moments, and CP violating EDM;
and tau physics at CLEO-III and at CLEO-c.
\vspace{1pc}
\end{abstract}

\maketitle

\section{Introduction}

Over the last dozen years, the CLEO Collaboration has
made use of data collected by the the CLEO-II detector \cite{ref:cleodet}
to measure many of the properties of the tau lepton and its neutrino.
Now that the experiment is making a transition from operation
in the 10 GeV (B factory) region to the 3-5 GeV tau-charm factory
region \cite{ref:cleoc},
it seemed to the author and the Tau02 conference organizers
to be a good time to review the contributions of 
the CLEO experiment to tau physics.
We are consciously omitting results from CLEO-I
(data taken before 1989).
The author has chosen to take a semi-critical approach,
emphasizing both the strengths and weaknesses of 
tau physics at CLEO: past, present, and future.


\section{Tau Production at 10 GeV}
\label{ss:production}

We begin by discussing a topic on which CLEO has {\it not} published:
tau pair production.

In $e^+e^-$ collisions, one can study taus in {\it production}
and/or {\it decay}. The production reaction
$e^+e^- \to \gamma^* \to \tau^+\tau^-$ is governed by
well-understood QED. As such, it is not terribly interesting
(or at least, nowhere near as interesting as studying
$e^+e^- \to \gamma^*/Z^{(*)} \to \tau^+\tau^-$ at LEP I and LEP II).
CLEO has studied resonance production and decay, \eg, in \cite{ref:Upstautau}
$e^+e^- \to \Upsilon(nS) \to \tau^+\tau^-$.

In addition to overall rate, one can search for small
anomalous couplings.
Rather generally, these can be parameterized as anomalous 
magnetic, and (CP violating) electric, dipole moments.
With $\tau^+\tau^-$ final states, 
one can study the spin structure of the final state; 
this is perhaps the most sensitive way to search for anomalous couplings.
We will return to this subject in section~\ref{ss:dipole}.

Measuring the {\it production} rate at 10 GeV
is only interesting if it is precise: $< 1\%$.
This has proven to be difficult at CLEO, for several reasons.

First, it is desirable to do an inclusive selection of $\tau^+\tau^-$
final states, so that production rates don't depend on 
decay branching fractions, which were not measured at the
1\%\ level in the early part of the last decade.
It's difficult to select taus inclusively,
because backgrounds from $q\bar{q}$, $e^+e^-$, $\mu^+\mu^-$ and two-photon
are less easily distinguishable from $\tau^+\tau^-$ than at LEP,
and they depend on tau decay mode.
It's not really hard, but it's hard to get {\it precise} 
selection efficiencies. In the end, tau selection at CLEO always
had smaller efficiencies and/or larger backgrounds, and often
also larger systematic errors, than analogous analyses at LEP.

To measure the production cross-section, one needs a
precise luminosity measurement, with an error $<< 1\%$.
At CLEO, we selected
large angle Bhabhas ($e^+e^-$), $\mu^+\mu^-$, $e^+e^-\to \gamma\gamma$
with rather high statistics.
To get the luminosity, we need accurate 
predictions of the cross-section times selection efficiency
from precision Monte Carlo simulations,
incorporating accurate QED radiative corrections.
Much less effort has gone into this at 10 GeV than at LEP!
Each of these measurements has small statistical errors,
but the systematic errors are of the order of $\sim 2\%$.
CLEO got agreement between the 3 QED processes at the level of 1\%
but not much better; the discrepancies are likely to be in the 
QED Monte Carlos. 
CLEO quotes a 1\%\ systematic error on luminosity \cite{ref:lumin}.

The moral is that further progress in this topic
requires {\it precision} QED Monte Carlos.
The {\bf KKMC} program \cite{ref:KKMC}
promises precision results,
but it must be validated with careful computational
and experimental cross-checks.
Consistent results for $e^+e^-$, $\mu^+\mu^-$, $e^+e^-\to \gamma\gamma$
are a good first step.
Further, measurements of anomalous moments make use of
the spin correlations in tau pair production at 10 GeV,
so one must validate the correct treatment of
spin-dependence in the Monte Carlo.

\section{Tau decay physics at CLEO}

CLEO pursued a systematic study of all tau decays
during the 1990's, learning much about the tau,
its neutrino, and low-energy meson dynamics.
The main decay modes are listed in Table~\ref{tab:modes}.

\begin{table}
\begin{center}
\caption{\label{tab:modes}
Main decay modes of the $\tau$ lepton.}
\begin{tabular}{l l l}
\hline
 $\tau\rightarrow e\nu\nu_\tau$          & $\approx 18\%  $  & 
Br, univ, Michel \\
 $\tau\rightarrow \mu\nu\nu_\tau$        & $\approx 17\%  $  & 
Br, univ, Michel \\
 $\tau\rightarrow \pi\nu , \, K\nu_\tau$ & $\approx 12\%  $  & 
Br, univ  \\
 $\tau\rightarrow \pi\pi\nu_\tau$        & $\approx 25\%  $  & 
Br, $\rho$, $\rho^\prime$, CVC, $\Pi$ \\ 
 $\tau\rightarrow K\pi\nu_\tau$          & $\approx 1.4\% $  & 
Br, $K^*$, $K^{*\prime}$ \\ 
 $\tau\rightarrow 3\pi\nu_\tau$          & $\approx 18\%  $  & 
Br, $a_1$, $a_1^\prime$, $\hnt$ \\ 
 $\tau\rightarrow K\pi\pi\nu_\tau$       & $\approx 0.8\% $  & 
Br, $K_1$, $K_{1b}$, W-Z \\
 $\tau\rightarrow 4\pi\nu_\tau$          & $\approx 5\%   $  & 
Br, $\rho^\prime$, CVC  \\
  $\tau\rightarrow \mbox{rare}$     & $\approx 2\%   $  & 
$5\pi$, $6\pi$, $\eta\pi\pi$, ... \\
  $\tau\to \eta\pi\nu, b_1\nu_\tau$       & $\ll 1\%     $  & 
2$^{nd}$-class currents \\
  $\tau\rightarrow \mbox{forbidden}$ & $\ll 1\%     $  & 
neutrinoless decays \\
\hline
\end{tabular}
\end{center}
\end{table}

\subsection{One-prong problem}

In the early 90's, the ``tau one-prong'' problem was raging; 
the branching fractions
for exclusively reconstructed tau decays
didn't add up to 1 \cite{ref:PDG94}.
The resolution of this discrepancy
required precision (sub-1\%) branching fractions measurements.

CLEO-II was a new detector; acceptances, in/efficiencies, 
and detector simulation needed to be understood very well.
Important issues included the 
detection of $\pi^0$'s: they rarely merged into one shower,
and CLEO obtained great $m_{\gamma\gamma}$ resolution ($\sim$ 6 MeV)
with its CsI calorimeter.
However,
soft photons could get lost; and
``splitoffs'' from hadronic showers could fake photons. 
Overall, ensuring reliable detection of
$\pi^0$'s resulted in a detection efficiency $\sim 50(1\pm 0.03)\%$.
Also, CLEO had poor K/$\pi$ separation over most of the
interesting momentum range. 
We made progress using $K^0_S$, with detection efficiency $\sim 50\%$.
Because of all this, it was
hard to know the overall detection efficiency, 
after backgrounds, to better than 1\%.

Ultimately, CLEO's branching fraction measurements 
were limited by knowledge of luminosity
(1\%), cross-section 
(computed to order $\alpha$ with KORALB \cite{ref:KORALB}, $\sim 1\%$),
and knowledge of the detection efficiency and backgrounds ($\sim 1-2\%$).
However, with millions of produced $\tau^+\tau^-$,
what we lacked in efficiency we made up for in statistics.

By 1995, we made $\sim (1-2)\%$ measurements of
branching fractions 
to $e\nu\nu_\tau$, $\mu\nu\nu_\tau$, $\pi/K\nu_\tau$, $\pi\pi^0\nu_\tau$, 
$\pi n\pi^0\nu_\tau$, 
$3\pi\nu_\tau$, $3\pi\pi^0\nu_\tau$, 
\etc \cite{ref:univ,ref:CLEOrhoBR,ref:npio,ref:3pi}.
These measurements 
reduced the ``tau one-prong'' problem to insignificance by 
PDG 1996 \cite{ref:PDG96}.
We had also
tested $e/\mu/\tau$ charged-current coupling universality at 
the 1\%\ level \cite{ref:univ}.
By then, LEP was measuring branching fractions with
total errors much smaller than $1\%$. 
This came as quite a shock to CLEO!
It helped that the LEP experiments 
knew $N_{\tau\tau} = \sigma{\cal L}$ quite well,
as a by-product of their incredibly successful electroweak program.


\section{Rare semi-hadronic decay modes}

Rare semi-hadronic decay modes of the tau provide unique
laboratories for low-energy meson dynamics,
and tests of conservation laws.
With the world's largest sample of tau pairs throughout the 1990's,
CLEO made many first and/or most precise
measurements of rare decay modes \cite{ref:3pi2pi0,ref:5pi,ref:3pi3pi0,ref:6pi,ref:7pi},
listed in Table~\ref{tab:Rmodes}.

\begin{table}
\begin{center}
\caption{\label{tab:Rmodes}
Measurements of branching fractions for rare semi-hadronic
$\tau$ decay modes by CLEO-II.}
\begin{eqnarray*}
\hline \\
  \BR(2\pi^-\pi^+2\pi^0 \nu_\tau)  &= (5.3\pm 0.4)\times 10^{-3} \\
  \BR(3\pi^-2\pi^+      \nu_\tau)  &= (7.8\pm 0.6)\times 10^{-4} \\
  \BR(2\pi^-\pi^+3\pi^0 \nu_\tau)  &= (2.2\pm 0.5)\times 10^{-4} \\
  \BR(3\pi^-2\pi^+\pi^0 \nu_\tau)  &= (1.7\pm 0.3)\times 10^{-4} \\
  \BR(\pi^-2\pi^0\omega \nu_\tau)  &= (1.5\pm 0.5)\times 10^{-4} \\
  \BR(2\pi^-\pi^+\omega \nu_\tau)  &= (1.2\pm 0.3)\times 10^{-4} \\
  \BR(3\pi^-2\pi^+2\pi^0\nu_\tau)  &< 1.1\times 10^{-4} \\
  \BR(7\pi^\pm(\pi^0)   \nu_\tau)  &< 2.4\times 10^{-6}. \\
\hline
\end{eqnarray*}
\end{center}
\end{table}

\begin{figure}[!ht]
\begin{center}
\psfig{figure=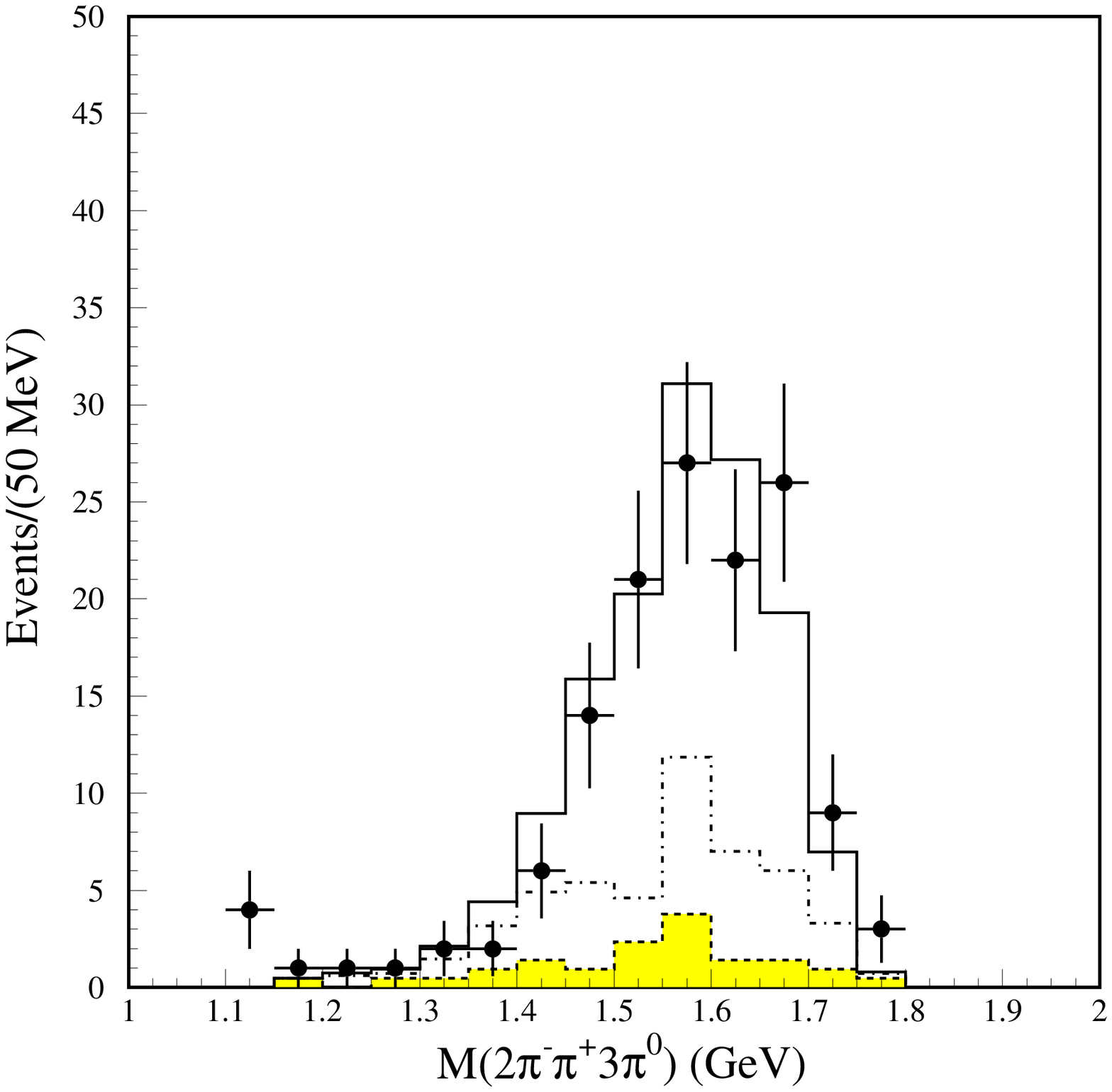,width=1.4in}
\psfig{figure=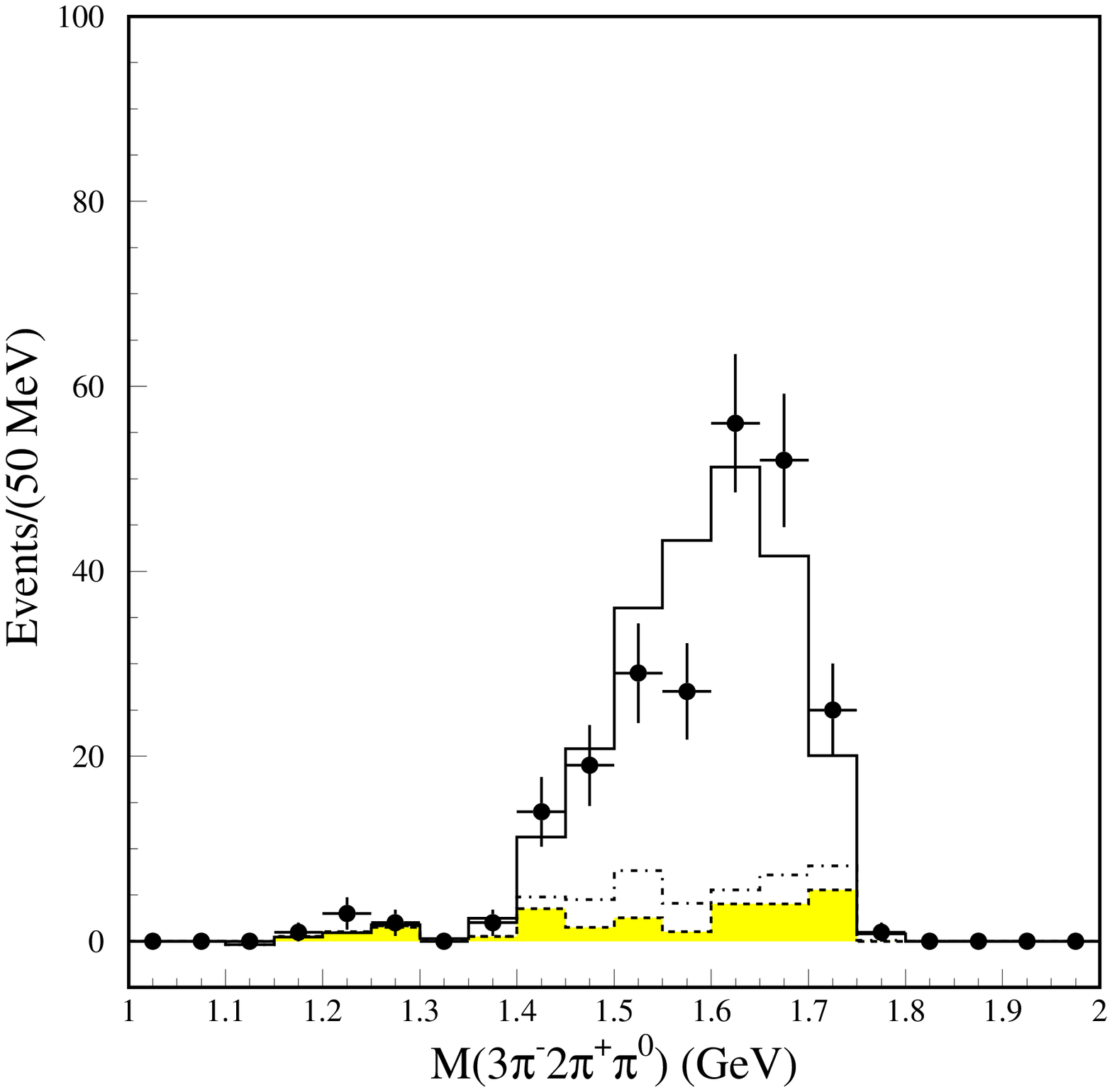,width=1.4in}
\caption{\label{fig:6pi}
Mass distributions from $\tau\to 6\pi\nu_\tau$.
The data are shown as data points
with error bars, the lower histograms
show various background models, and 
the open solid histogram shows the model of the 
background plus tau decay signal.
Left: $M(2\pi^-\pi^+ 3\pi^0\nu_\tau)$ in
 $\tau^-\to 2\pi^-\pi^+ 3\pi^0\nu_\tau$ \protect\cite{ref:3pi3pi0}.
Right: $M(3\pi^-2\pi^+ \pi^0)$ in
$\tau^-\to 3\pi^-2\pi^+ \pi^0\nu_\tau$ \protect\cite{ref:6pi}.
}
\end{center}
\end{figure}

Decay modes like 
$5\pi$, $6\pi$, $7\pi$, $\eta\pi\pi$, $\eta3\pi$, 
are relatively easy to reconstruct.
The big problem is background from $q\bar{q}$.
Lepton tags can clean that up reasonably well, 
but an irreducible background remains,
that can be estimated and subtracted statistically.
These modes have 
rich and complicated sub-structure,
which we attempted to delve into for the $5\pi$, $6\pi$, 
and $\eta3\pi$ modes.
Some $6\pi$ signals are shown in Fig.~\ref{fig:6pi}.

Of particular interest are modes involving $\eta$ mesons
\cite{ref:etapi,ref:etaK,ref:eta3pi,ref:etaKstar},
listed in Table~\ref{tab:etamodes}.

\begin{table}
\begin{center}
\caption{\label{tab:etamodes}
Measurements of 
$\tau$ decay modes involving kaons from CLEO-II.}
\begin{eqnarray*}
\hline \\
  \BR(\nu_\tau \eta\pi^-) &<& 1.4\times 10^{-4} \mbox{at 95\%\ CL}\\
  \BR(\nu_\tau \eta K^-) &=& (2.6\pm 0.5)\times 10^{-4} \\
   \BR(\nu_\tau \eta\pi^-\pi^0) &=& (1.7\pm 0.3)\times 10^{-3} \\
   \BR(\nu_\tau \eta\pi^-\pi^+\pi^-) &=& (3.4\pm 0.8)\times 10^{-4} \\
   \BR(\nu_\tau \eta\pi^-\pi^0\pi^0) &=& (1.4\pm 0.6)\times 10^{-4} \\
   \BR(\nu_\tau K^{*-}\eta) &=& (2.90\pm 0.80\pm 0.42)\times 10^{-4}. \\
\hline
\end{eqnarray*}
\end{center}
\end{table}

\begin{figure}[!ht]
\begin{center}
\psfig{figure=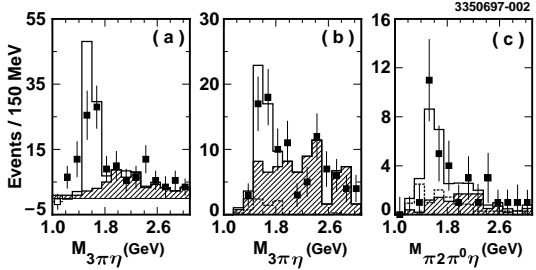,width=2.8in}
\caption{\label{fig:3pieta}
Mass distributions from $\tau\to 3\pi\eta\nu_\tau$ \protect\cite{ref:eta3pi}.
The data are shown as data points
with error bars, the hatched histogram
shows the background model (mostly from non-tau continuum), 
and the open solid histogram shows the model of the 
background plus tau decay signal.
$M((3\pi)^-\eta)$ is plotted, for 
(a) $2\pi^-\pi^+\eta$, $\eta\to\gamma\gamma$;
(b) $2\pi^-\pi^+\eta$, $\eta\to 3\pi^0$;
(c) $\pi^- 2\pi^0\eta$, $\eta\to\gamma\gamma$.
}
\end{center}
\end{figure}

The $\eta\pi^-$ mode is a signature for second-class 
(isospin-violating) currents.
The $\eta K^-$ mode proceeds by SU(3)$_f$ violation.
The $\eta\pi^-\pi^0$ mode proceeds by the Wess-Zumino 
anomalous current.
The $\eta3\pi$ signals, which are rich in sub-structure,
are shown in Fig.~\ref{fig:3pieta}.

CLEO also observed the radiative 
decay modes $\tau^-\to e^- \nu_\tau\gamma$
and $\tau^-\to \mu^- \nu_\tau\gamma$ \cite{ref:lnugamma},
and made the first observation of the decays
$\tau^-\to e^- e^+e^-\bar{\nu}_e \nu_\tau$ (5 events) 
and $\tau^-\to \mu^- e^+e^-\bar{\nu}_\mu\nu_\tau$ (1 event) \cite{ref:3lnunu}.

\subsection{Rare strange modes - $X_S^-\nu_\tau$}

Modes with $K^0_S$ are easily accessible in the CLEO data.
Identifying charged kaons was problematical in CLEO-II,
but even with poor K/$\pi$ separation ($\sim < 2 \sigma$),
we identified and measured branching fractions for
many modes containing kaons, listed in Table~\ref{tab:Kmodes}.
During the same time period,
LEP-I produced terrific results on modes with kaons, 
including $K^0_L$!

\begin{table}
\begin{center}
\caption{\label{tab:Kmodes}
Measurements of 
$\tau$ decay modes involving kaons from CLEO-II.}
\begin{tabular}{ll}
\hline
$\BR(\bar{K^0}\pi^-\pi^0\nu_\tau)$    & = ($0.417\pm 0.058\pm 0.044$)\%\\
$\BR( K^-\pi^+\pi^-\nu_{\tau})$       & = ($0.345\pm 0.023 \pm 0.055$)\%\\
$\BR( K^-\pi^0\pi^0\nu_\tau)$         & = ($0.14\pm 0.10\pm 0.03$)\%\\
$\BR( K^-K^0\pi^0\nu_\tau)$           & = ($0.145\pm 0.036\pm 0.020$ )\%\\ 
$\BR( K^-K^+\pi^-\nu_{\tau})$         & = ($0.144\pm 0.013 \pm 0.028$ )\%\\
$\BR( K^0_S K^0_S\pi^-\nu_\tau)$      & = ($0.023\pm 0.005\pm 0.003$ )\%\\
$\BR( K^-\pi^+\pi^-\pi^0\nu_{\tau})$  & = ($0.075\pm 0.026 \pm 0.017$)\%\\
$\BR( K^-K^+\pi^-\pi^0\nu_{\tau})$    & = ($0.033\pm 0.018 \pm 0.007$)\%\\
\hline
\end{tabular}
\end{center}
\end{table}

\section{Forbidden (neutrinoless) decays}

Tau decays to final states with no $\nu_\tau$
(or more precisely, no missing energy) are a clear signature
for lepton-number violation, pointing directly to physics
beyond the Standard Model.
The golden mode is $\tau\to\mu\gamma$,
which in many models, would be the easiest to observe
in tau decays (despite the strong constraint from
non-observation of $\mu\to e\gamma$).

\begin{figure}[!ht]
\begin{center}
\psfig{figure=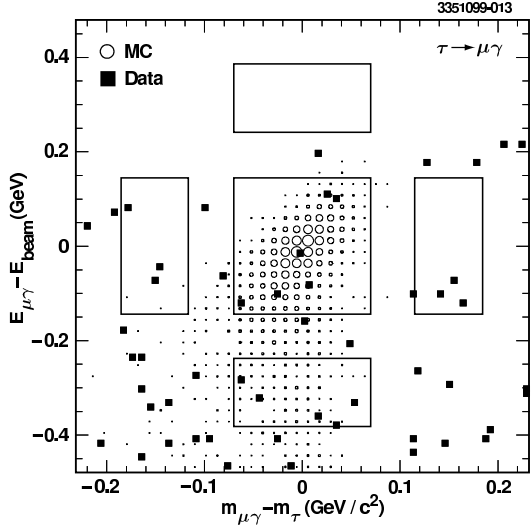,height=7cm}
\caption{\label{fig:mugamma}
CLEO data (points) and Monte Carlo prediction (open boxes)
for $\tau^-\to\mu^-\gamma$ \protect\cite{ref:mugamma99}.}
\end{center}
\end{figure}

CLEO-II set its first upper limit
$\BR(\tau\to\mu\gamma) < 4.2\times 10^{-6}$ (90\%\ CL)
in 1992 \cite{ref:mugamma92}.
This was 
improved to $3.0\times 10^{-6}$ by 1996 ($4.3\times 10^6$ tau pairs)
\cite{ref:mugamma97};
then $1.1\times 10^{-6}$ by 1999 
($12.6\times 10^6$ tau pairs) \cite{ref:mugamma99}.
By this point, we were starting to hit background events
(see Fig.~\ref{fig:mugamma}),
which appear to be irreducible; further progress can no longer
be expected to scale inversely with the number of produced
tau pairs.

CLEO also searched for many other neutrinoless modes,
setting limits on 22 of them in 1994 \cite{ref:nless94},
with branching fraction upper limits around $10^{-5}$.
This was updated in 1997 \cite{ref:nless97},
with 28 modes, including resonances;
branching fraction upper limits of a few $\times 10^{-6}$
were obtained.
We added 10 more modes with $\pi^0$'s and/or $\eta$'s 
in 1997 \cite{ref:nless97b}.
Five more modes were added in 1998 \cite{ref:nless98},
containing (anti-)protons: $\tau^-\to \bar{p} X^0$.

Throughout all this, we managed to forget about
modes containing $K^0$'s! 
This has now been corrected:
New for this conference \cite{ref:nlessK0}, based on 
$12.7\times 10^6$ $\tau^+\tau^-$), 
are the branching fraction upper limits at 90\%\ CL
listed in Table~\ref{tab:K0modes}.

\begin{table}
\begin{center}
\caption{\label{tab:K0modes}
Measurements of neutrinoless
$\tau$ decay modes involving $K^0_S$'s, from CLEO-II \protect\cite{ref:nlessK0}.}
\begin{eqnarray*}
\hline \\
\BR(e^- K^0_S)         &<& 9.1\times 10^{-7} \\
\BR(\mu^- K^0_S)       &<& 9.5\times 10^{-7} \\
\BR(e^- K^0_S K^0_S)   &<& 2.2\times 10^{-6} \\
\BR(\mu^- K^0_S K^0_S) &<& 3.4\times 10^{-6} \\
\hline
\end{eqnarray*}
\end{center}
\end{table}

\section{CLEO Michel Parameter analyses}

We turn now from measurements of branching fractions
to the {\it substructure} in the multi-particle decay modes.
For the leptonic decays $\tau^-\to \ell^- \bar{\nu}_\ell \nu_\tau$,
this amounts to the measurement of the Michel parameters
governing the Lorentz structure of the decay
(\ie, the search for deviations from the Standard Model $V-A$ structure).
Information on the Lorentz structure of the decay,
especially on the helicity of the tau neutrino $\hnt$,
can also be obtained from decay distributions
in semi-hadronic ($\tau^-\to X_h^- \nu_\tau$) decays.

CLEO-II has published four different analyses 
focusing on Lorentz structure:

\begin{itemize}
\item Select {$\ell^-\nubar\nutau$} 
      \vs\ {$\pi^{+}\pizero\nutaubar$};
      Use {$\pi^\pm\pi^0$} as tag. 
      From the lepton energy spectrum,
      extract the Michel parameters {$\rho$} and {$\eta$}
      \cite{ref:michel1}.

\item Select {$\pi^-\nutau$}
      \vs\ {$\pi^+\nutaubar$}; 
      exploit the spin correlations between the
      two taus in the event to extract
      the square of the tau neutrino helicity $|\hnt|^2$
      \cite{ref:hnt}.

\item Select {$\ell^-\nubar\nutau$}
      \vs\ {$\pi^{+}\pizero\nutaubar$};
      use the $\pi^\pm\pi^0$ decay as spin analyzer. 
      Use the full event kinematics to extract
      measurements of the Michel parameters
      {$\rho$}, {$\eta$}, {$\xi$}, 
      {$\delta$}, and {$|\hnt|$}
      \cite{ref:michel2}.

\item Select {$\ell^-\nubar\nutau$}
      \vs\ {$(\rho\pi)^-\nutaubar$}.
      Exploit the interference between the two $\rho\pi$ amplitudes,
      and use the full event kinematics 
      to extract the parity-violating {\it signed} tau neutrino helicity
      $\hnt$
      \cite{ref:a1model}.

\end{itemize}

\begin{figure}[!ht]
\begin{center}
\psfig{figure=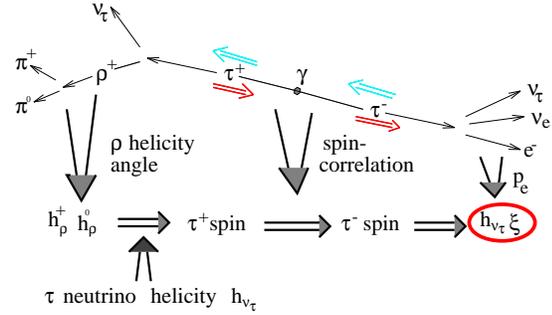,width=2.8in}
\caption{\label{fig:spincor}
Illustration of how to extract spin-dependent
Michel parameters in the reaction
$e^+e^-\to \tau^+\tau^-$,
$\tau^+\to\rho^+\nu_\tau$, $\tau^-\to\ell^-\nubar\nutau$.}
\end{center}
\end{figure}

The third-listed analysis \cite{ref:michel2},
in particular, made rather precise measurements, using a 
powerful technique.
To make full use of kinematical information, 
a full multi-dimensional likelihood fit was performed.
The fit correlated information on the $\rho^+$ polarization in
$\tau^+\to\rho^+\nu_\tau$ from the decay $\rho^+\to\pi^+\pi^0$,
the QED-predicted correlations between the spins of the
$\tau^+$ and the $\tau^-$, and the momentum distribution 
of the daughter lepton in $\tau^-\to\ell^-\nubar\nutau$,
in order to extract the spin-independent 
Michel parameters $\rho$ and $\eta$, the spin-dependent
parameters $\xi$ and $\xi\delta$, and 
the tau neutrino helicity $\hnt$.
This is illustrated in Fig.~\ref{fig:spincor}.

\begin{figure}[!ht]
\centerline{
\psfig{figure=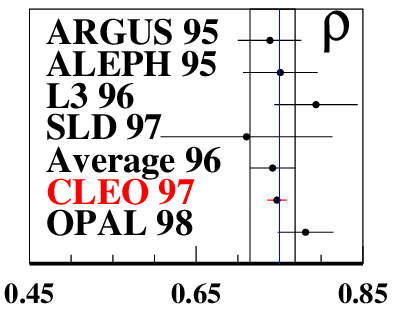,width=1.4in}
\psfig{figure=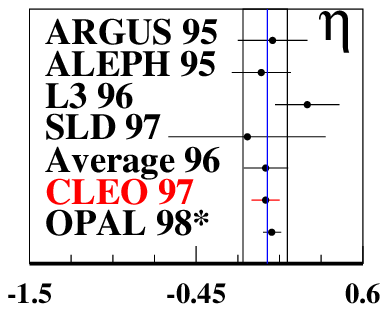,width=1.4in}
}
\centerline{
\psfig{figure=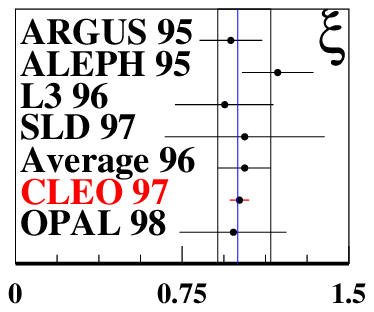,width=1.4in}
\psfig{figure=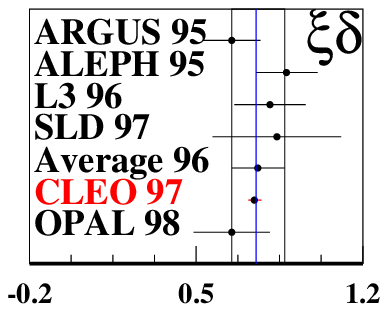,width=1.4in}
}
\centerline{
\psfig{figure=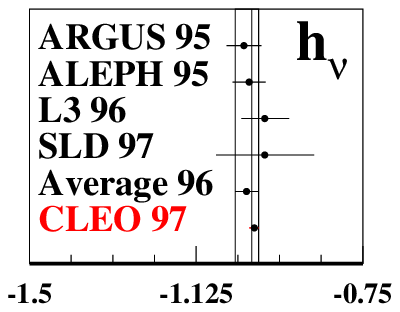,width=1.4in}
}
\caption{\label{fig:michelmeas}
Measurements of tau Michel parameters from CLEO'97 and
from other experiments.
$^*$The leptonic branching fractions were used to further
constrain the value of  $\eta$ in the OPAL 98 measurement.}
\end{figure}

The Michel parameters measured in this way
are compared with those from other experiments
in Fig.~\ref{fig:michelmeas}.
From these measurements, strong constraints could be placed on 
right-handed $\tau - \nu_\tau$ couplings and the mass
of a right-handed $W^\pm_R$.
Further, the precise limit obtained on $\eta$
(without making use of any constraint on $\eta$
from the leptonic branching fractions)
sets a constraint on the presence of, \eg,
a scalar charged Higgs mediating tau decays.

\section{Hadronic substructure in tau decays}

Tau semi-hadronic decays $\tau^- \to X_h^- \nu_\tau$
provide a uniquely clean probe of low energy meson dynamics.
Recall that strong dynamics is the most poorly understood
part of the Standard Model.
The fundamental theory is QCD, but it is difficult to use QCD
to characterize hadronic structure in detail.
Instead, we must rely on models, symmetries and conservation laws
(such as isospin and $SU(3)_f$),
the Conserved Vector Current (CVC) hypothesis, sum rules,
chiral perturbation theory, and results from QCD on the lattice.

The momentum transfer is small in $\tau$ decays;
we are in the region where resonances dominate,
and their description relies on phenomenological models.
The dynamics of such hadronic systems is 
parameterized in tau decays via
the ``spectral function'' $v(q^2)$ ($q^2 = M(X_h)^2$), 
containing all the strong dynamics. 
These spectral functions can be related, via CVC, to similar 
quantities in $e^+e^-$ collisions and in vacuum polarization
diagrams that arise in $(g-2)_\mu$ and $\alpha(q^2)$ \cite{ref:davier}.
CLEO can measure $v(q^2)$ using exclusive final states;
it is more problematic to do inclusive studies
such as has been done at LEP, as discussed above.
Still, CLEO published one paper on the subject \cite{ref:alphas},
extracting $\alpha_S(m^2_\tau)$.


CLEO has studied substructure in the following tau decays:

\begin{itemize}
\item $\tau^- \to \pi^-\pi^0\nu_\tau$ \cite{ref:CLEOrho};
      see Fig.~\ref{fig:pimodes}.
      We extracted the 
      mass and width of the $\rho^-$ meson and
      the mass, width, and coupling of the $\rho^{\prime-}$ meson;
      measured the pion form factor $|F_\pi(q^2)|$; 
      and provided tests of CVC in comparison with $e^+e^-\to\pi^+\pi^-$
      (see \cite{ref:eidelman}).
      These data are useful in the evaluation of 
      the hadronic contribution to the vacuum polarization
      diagrams that arise in $(g-2)_\mu$ and $\alpha(q^2)$
      \cite{ref:davier}.
\item $\tau^- \to \pi^- K^0\nu_\tau$ \cite{ref:CLEOKstar},
      see Fig.~\ref{fig:Kmodes}.
      We studied the mass and width of $K^{*-}$ meson, 
      looked for evidence of a $K^{*\prime-}$,
      and extracted the decay constant $f_{K^*}$.
      This work has not yet been published!
      Of particular note is a significant discrepancy 
      between the mass of the $K^{*-}$ seen in our
      clean sample from $\tau^- \to K^{*-}\nu_\tau$
      and the world average in the PDG.
\item $\tau^- \to 3\pi\nu_\tau$;
      see Fig.~\ref{fig:pimodes}.
      We have performed three analyses:
      a model-dependent analysis using the $\pi^-\pi^0\pi^0$ mode
      \cite{ref:a1model},
      a model-independent measurement of the structure functions
      using the $\pi^-\pi^0\pi^0$ mode \cite{ref:a1spec},
      and a model-dependent analysis using the $\pi^-\pi^+\pi^-$ mode
      \cite{ref:shibata}.
      We found a rich structure, including the presence
      of scalar and tensor mesons.
      We measured the decay constant $f_{a_1}$;
      the signed neutrino helicity $h_{\nu_\tau}$;
      and limits on couplings to the $\pi^\prime(1300)$.
\item $\tau^- \to (K\pi\pi)^-\nu_\tau$  
         \cite{ref:CLEOK0,ref:krav1,ref:krav2,ref:CLEOKstar};
      see Fig.~\ref{fig:Kmodes}.
      Here again, a rich structure can be uncovered.
      We have studied the mass, width, couplings
      and decay branching fractions of the $K_1(1270)$ 
      and $K_1(1400)$ mesons, and explored their 
      mixing and SU(3) violating couplings
      to the tau. More on this in section~\ref{ss:Kpipi}.

\item $\tau^- \to (4\pi)^-\nu_\tau$ \cite{ref:CLEO4pi}.
      The physics that can be explored in this decay
      is discussed in section~\ref{ss:4pi}.

\item $\tau^- \to \eta(3\pi)^-\nu_\tau$ \cite{ref:eta3pi}; 
      see Fig.~\ref{fig:3pieta}.
      This decay was first observed by CLEO,
      using two different modes of the $\eta$, 
      and two different charge combinations of $(3\pi)^-$.
      We observed clear evidence for $\tau^- \to f_1(1285)\pi^-\nu_\tau$,
      $f_1\to a_0(980) \pi$, $a_0\to \eta\pi$
      and measured their product branching fractions.
      We searched for and set upper limits on
      $\tau^- \to \eta^\prime \pi^-\nu_\tau$ and
      $\tau^- \to \eta^\prime \pi^-\pi^0\nu_\tau$.

\item $\tau^- \to 2\pi^-\pi^+3\pi^0\nu_\tau$ \cite{ref:6pi};
      see Fig.~\ref{fig:6pi}.
      We measured a branching fraction in good agreement
      with expectations from isospin, 
      and constrained the isospin structure of the decay.
      The branching fraction is somewhat below
      the CVC prediction from $e^+e^-$ data \cite{ref:eidelman}.
      The decay appears to be saturated by the channels
      $\tau^- \to \pi^-2\pi^0\omega\nu_\tau$,
      $\tau^- \to 2\pi^-\pi^+\eta\nu_\tau$, and
      $\tau^- \to \pi^-2\pi^0\eta\nu_\tau$.

\ei

Notable omissions from this list include a study of substructure in
$\tau^- \to (5\pi)^-\nu_\tau$,
and the study of Lorentz structure in $\tau^- \to \eta\pi^-\pi^0\nu_\tau$
(which is expected to proceed via the Wess-Zumino anomalous current).

\begin{figure}[!ht]
\begin{center}
\psfig{figure=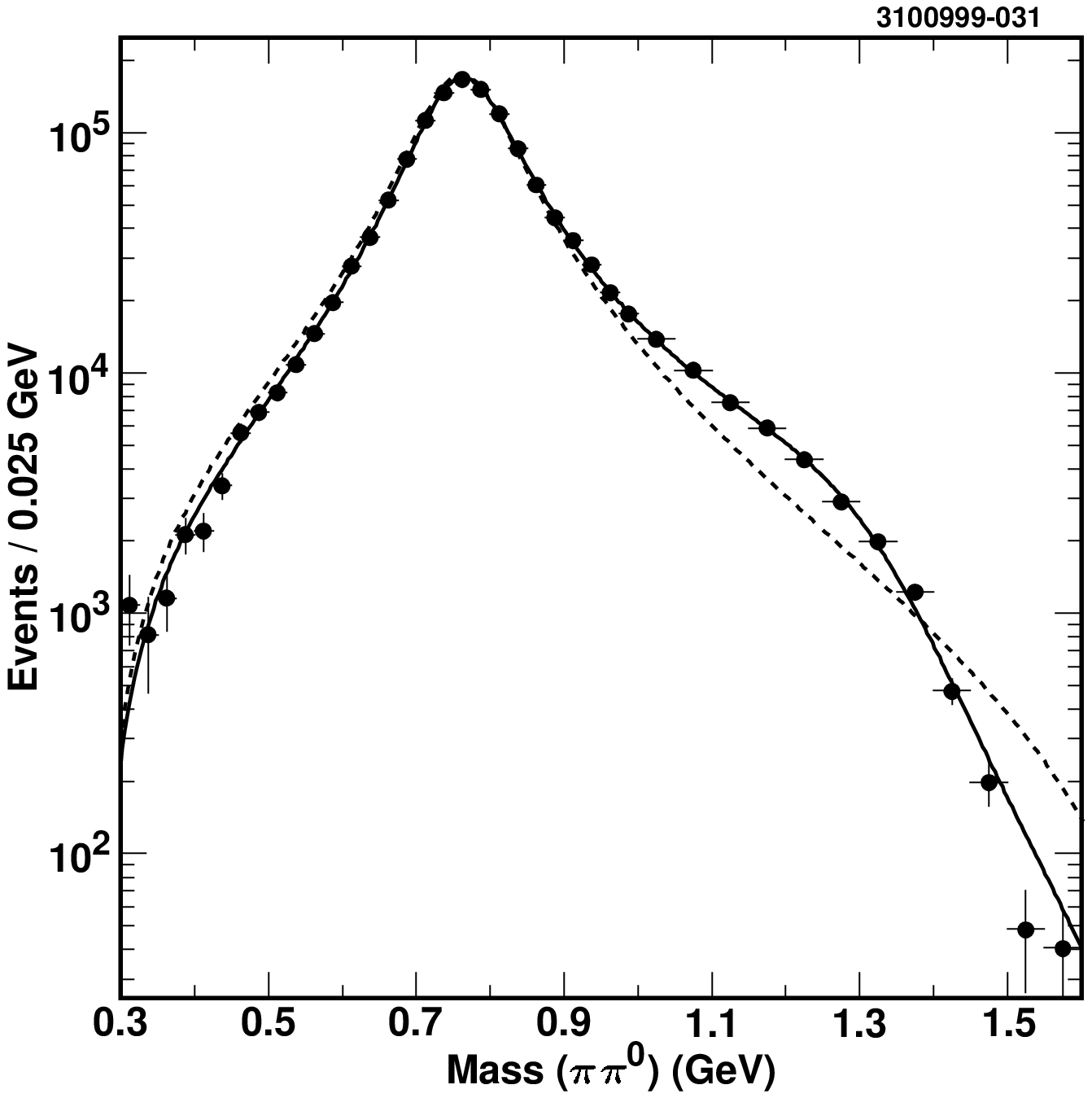,width=2.8in}
\psfig{figure=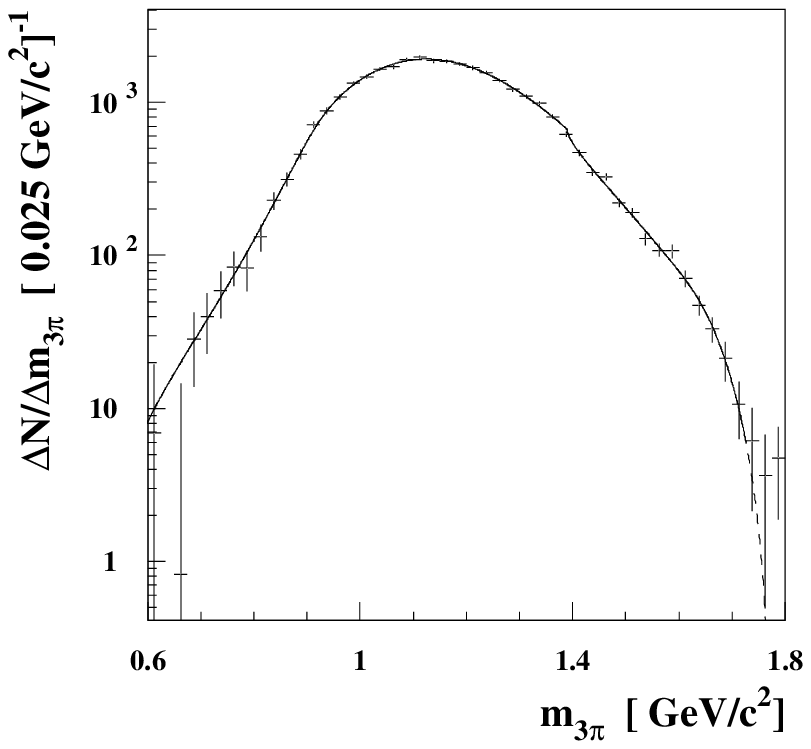,width=2.8in}
\caption{\label{fig:pimodes}
Fully corrected mass distributions from 
(top) $\tau^-\to \pi^-\pi^0\nu_\tau$ \protect\cite{ref:CLEOrho};
(bottom) $\tau^-\to \pi^-\pi^0\pi^0\nu_\tau$ \protect\cite{ref:a1model}.
The data are shown as data points,
and the curve
shows a model of the tau decay signal.
}
\end{center}
\end{figure}

\begin{figure}[!ht]
\begin{center}
\psfig{figure=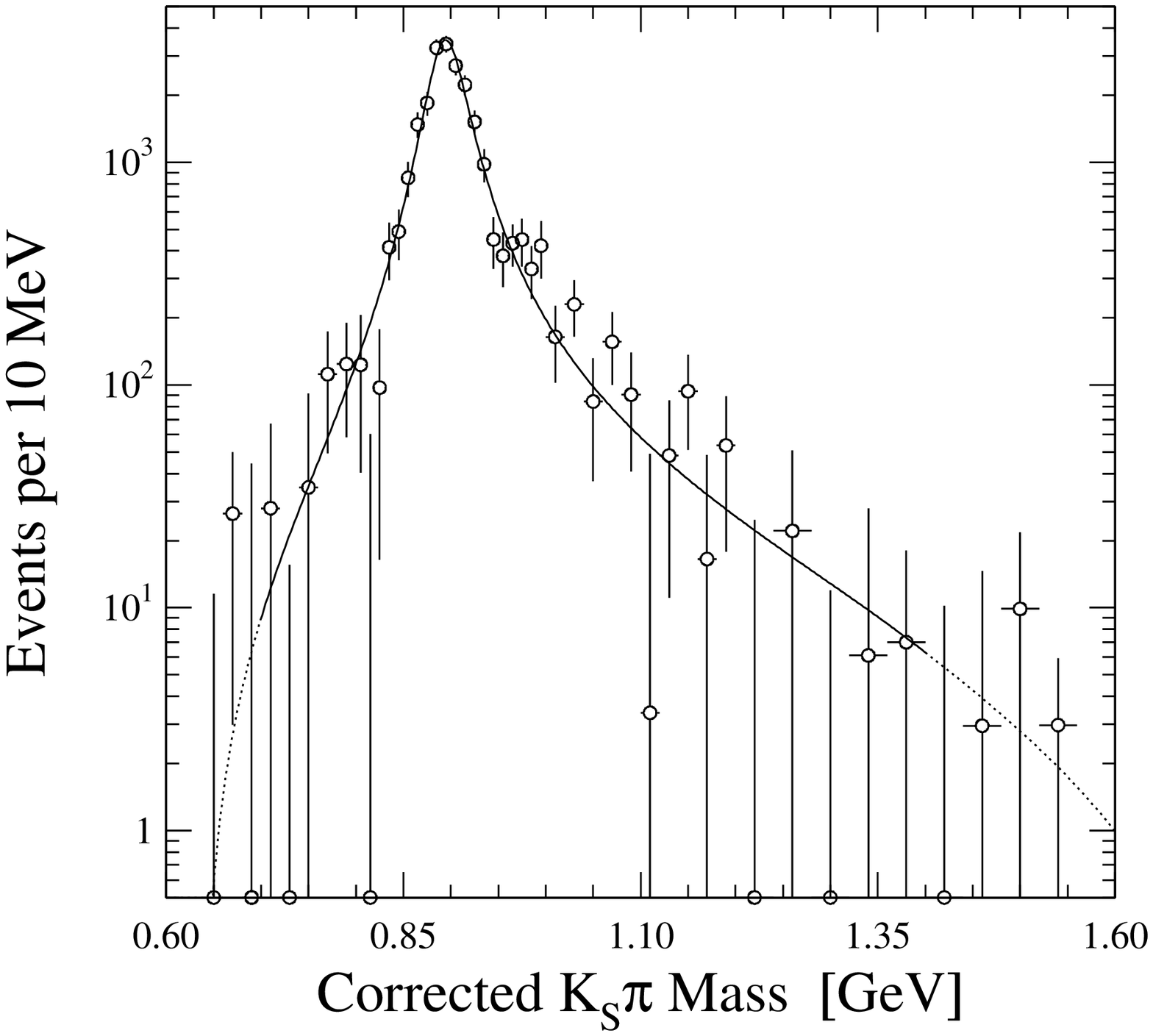,width=1.95in}
\psfig{figure=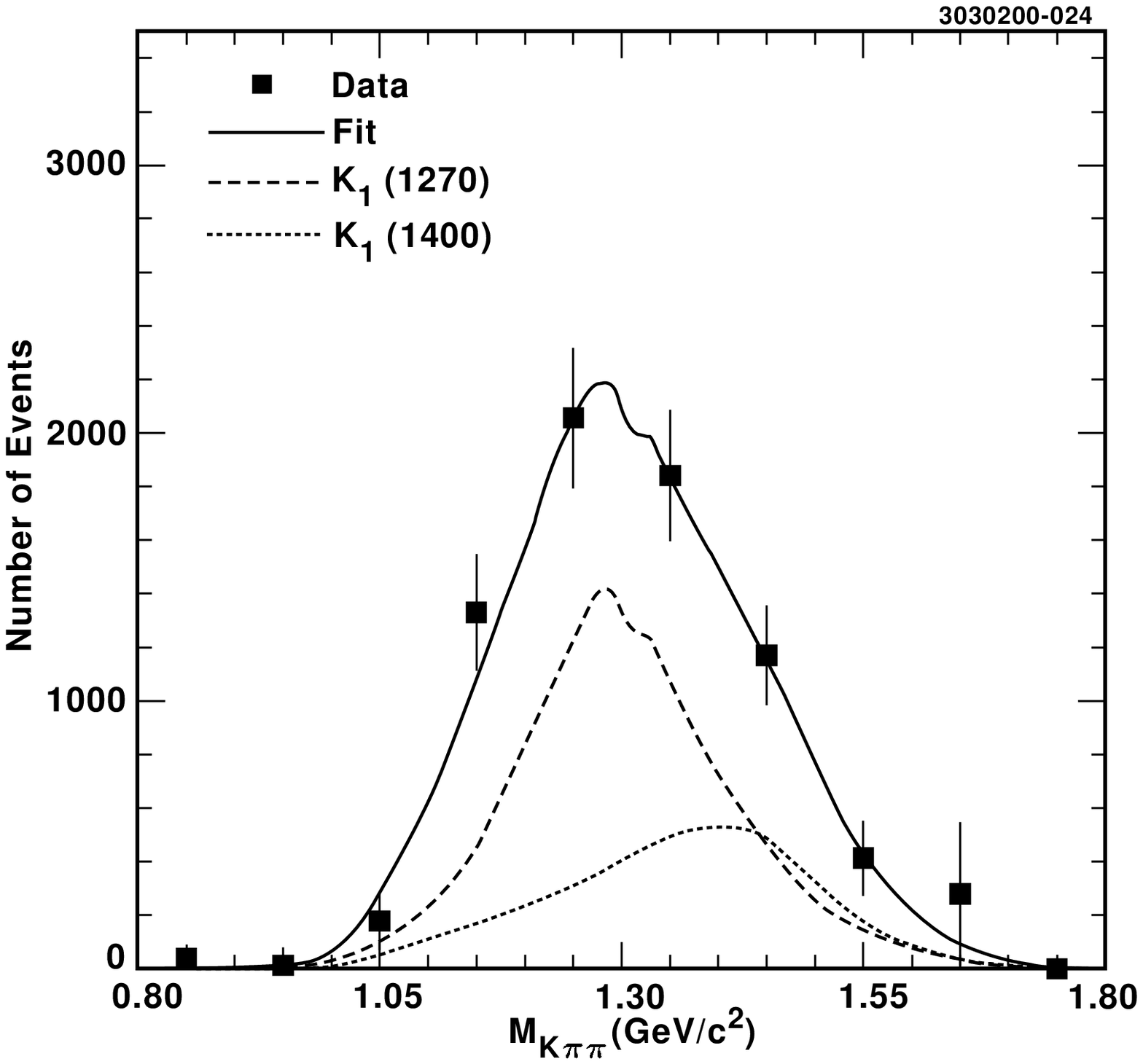,width=1.95in}
\psfig{figure=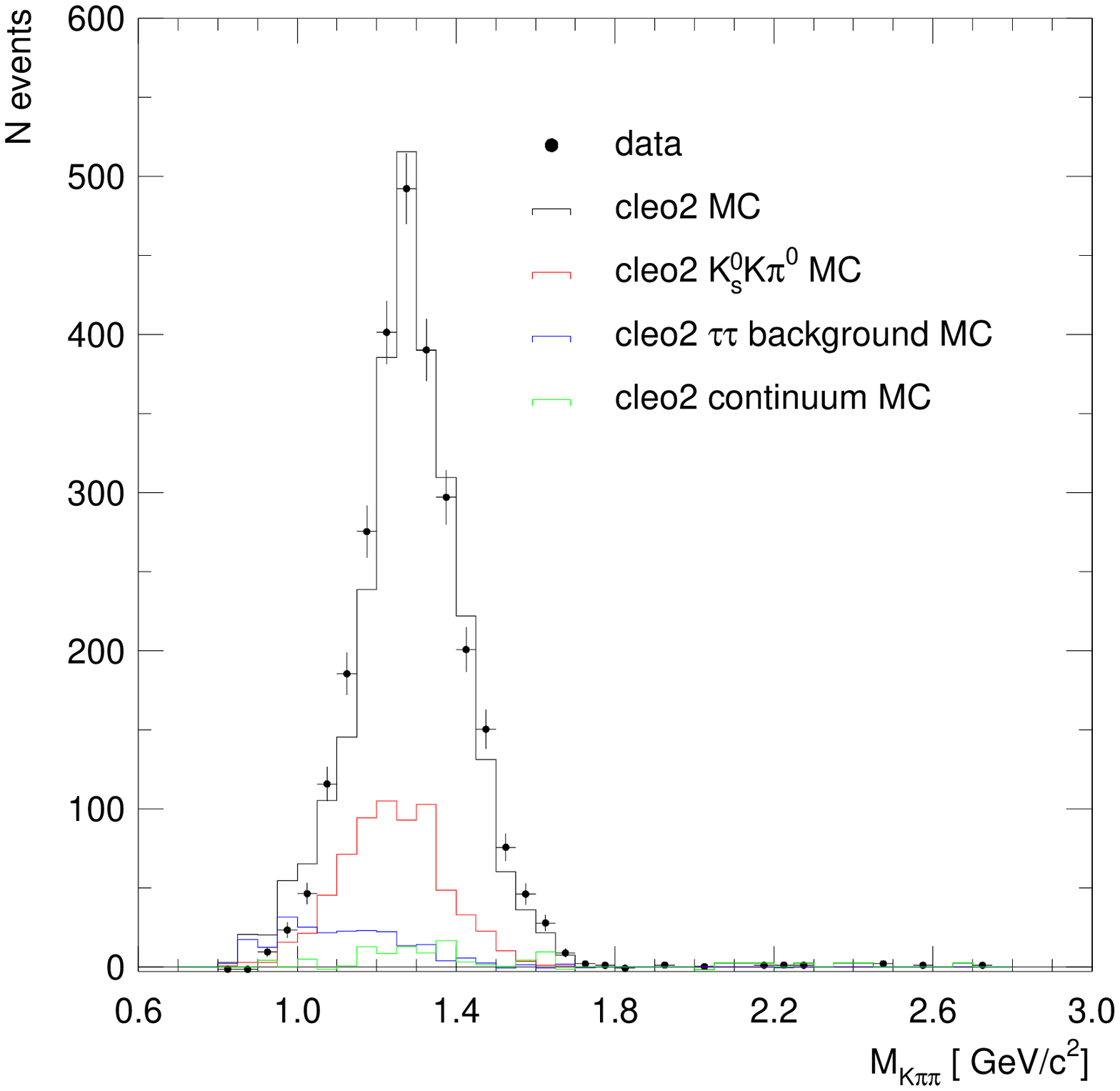,width=2.1in}
\caption{\label{fig:Kmodes}
Mass distributions from tau decay modes containing kaons.
Top: $M(K^0_S\pi^-)$ from $\tau^-\to K^0_S\pi^-\nu_\tau$ \protect\cite{ref:CLEOKstar}.
Middle: $M(K\pi\pi)$ from $\tau^-\to K^-\pi^+\pi^+\nu_\tau$ \protect\cite{ref:krav2}.
Bottom: $M(K\pi\pi)$ from $\tau^-\to K^0_S\pi^-\pi^0\nu_\tau$ \protect\cite{ref:CLEOKstar}.
The data are shown as data points,
and the topmost curve or histogram 
shows a model of the tau decay signal.
}
\end{center}
\end{figure}

\subsection{$\tau^- \to K_1^- \nu_\tau$ structure}
\label{ss:Kpipi}

The decay $\tau^- \to (K\pi\pi)^- \nu_\tau$
has particularly interesting structure.
It can be studied in the final states 
$K^-\pi^+\pi^-\nu_\tau$ and
$K^0_S\pi^-\pi^0\nu_\tau$.
CLEO has published results focusing on the former \cite{ref:krav2},
despite the poor $K^\pm/\pi^\pm$ separation.
Results on the latter mode, which is much cleaner
and can be isolated event-by-event, 
are still in progress \cite{ref:CLEOKstar}.

The $(K\pi\pi)^-$ final state is expected to be dominated by
the axial-vector $K_1$ ($J^P = 1^+$).
There are two such states; in the quark model, they are
$K_a$ in the $^3P_1$ octet, the strange partner of the $a_1(1260)$;
and $K_b$ in the $^1P_1$ octet, the strange partner of the $b_1(1235)$.
The $K_b$ couples to $W$ only through an 
SU(3)-violating ``second-class'' current.
Both $K_1$'s decay to $K\pi\pi$ via $K^*\pi$ and $K\rho$, 
and they mix, via virtual states, into 
the physically observed $K_1(1270)$, $K_1(1400)$.
So, we have weak coupling, SU(3)-violation, mixing,
and decays to resonances.
In addition, $(K\pi\pi)^-$ final state could proceed via
a vector current, via the Wess-Zumino anomaly: 
$K^{*\prime} \to (K^*\pi, K\rho) \to K\pi\pi$ .



We can parameterize the 
$K_{1a} \leftrightarrow K_{1b}$ mixing via a mixing angle:
\begin{eqnarray*}
 K_1(1400) &=& K_a \cos{\theta_K} - K_b \sin{\theta_K} \\
 K_1(1270) &=& K_a \sin{\theta_K} + K_b \cos{\theta_K}
\end{eqnarray*}
and the $SU(3)_f$ symmetry breaking via a parameter $\delta$: \\
$ \tau \to W \to |K_a\rangle - \delta|K_b\rangle $ \\
$ |\delta| = (m_s-m_u)/\sqrt{2}(m_s + m_u)\approx 0.18 $. \\
The branching fraction of the $\tau$ to the $K_1$'s
can then be written in terms of these parameters:
$$
\frac{{\cal B}(\tau\to K_1(1270)\nu)}
{{\cal B} (\tau\to K_1(1400)\nu)} = \left|
     \frac{\sin{\theta_K} - \delta\cos{\theta_K}}
     {\cos{\theta_K} + \delta\sin{\theta_K}}
               \right|^2 \times \Phi,
$$
where $\Phi$ is some known (or estimatable)
kinematical and phase space terms.

From the CLEO data \cite{ref:krav2},
we obtain two possible solutions,
depending upon the sign of $\delta$:
\begin{eqnarray*}
 \theta_K & = (69\pm 16\pm 19)^\circ ~~~(\delta=0.18), \\
 \theta_K & = (49\pm 16\pm 19)^\circ ~~~(\delta=-0.18).
\end{eqnarray*}
These mixing angles are consistent with those obtained
using only the $K_1$ widths and branching fractions \cite{ref:suzuki}.

\subsection{ $\tau^-\to (4\pi)^- \nu_\tau$ }
\label{ss:4pi}

The $\tau\to 4\pi\nu_\tau$ decay is expected to proceed 
through the vector current $(J^P=1^-)$,
dominated by the $\rho$, $\rho'$, $\rho''$... resonances.
There are many sub-resonances that can 
contribute: $\omega\pi$, $\eta\pi$, $a_1\pi$.
In the CLEO analysis \cite{ref:CLEO4pi},
the spectral functions $v(m^2_{4\pi})$
and $v(m^2_{\omega\pi})$ were measured,
and the $\omega\pi$ contribution was modeled 
with interfering $\rho$, $\rho'$, $\rho''$... resonances
(see Fig.~\ref{fig:4pi}).
The resonant substructure was measured and modeled,
showing clear evidence for $\omega\pi$
and $\rho\pi\pi$, and the latter is consistent
with originating from $a_1\pi$, $a_1\to \rho\pi$.

The $4\pi$ spectral function must be known well
in order to use this final state to kinematically 
constrain the $\nu_\tau$ mass from 
$\tau\to 4\pi\nu_\tau$ data; this is discussed in
section~\ref{ss:mnutau}.
The $4\pi$ spectral function can also be compared 
with the isospin-rotated reactions
$e^+e^-\to 2\pi^+2\pi^-$, $\pi^+\pi^-2\pi^0$
as a test of CVC \cite{ref:eidelman}.

The $4\pi$ final state can also proceed through
the axial-vector current, which is expected to be
dominated by the $b_1(1235)$:
$\tau\to b_1\nu_\tau $, $b_1\to \omega\pi$.
The $b_1$ has, however, the wrong G-parity 
to couple to the weak charged current; it is a second-class current.
The resulting decay to $\omega\pi$ occurs via an
S- or D-wave, instead of the $\rho\to\omega\pi$ P-wave.
CLEO measured the angular distribution 
in this decay 
(Fig.~\ref{fig:4pi})
and found complete consistency
with P-wave decay, setting a limit on 
the non-vector current contribution of less than 5.4\%\ of the total
at 90\%\ CL.

\begin{figure}[!ht]
\begin{center}
\psfig{figure=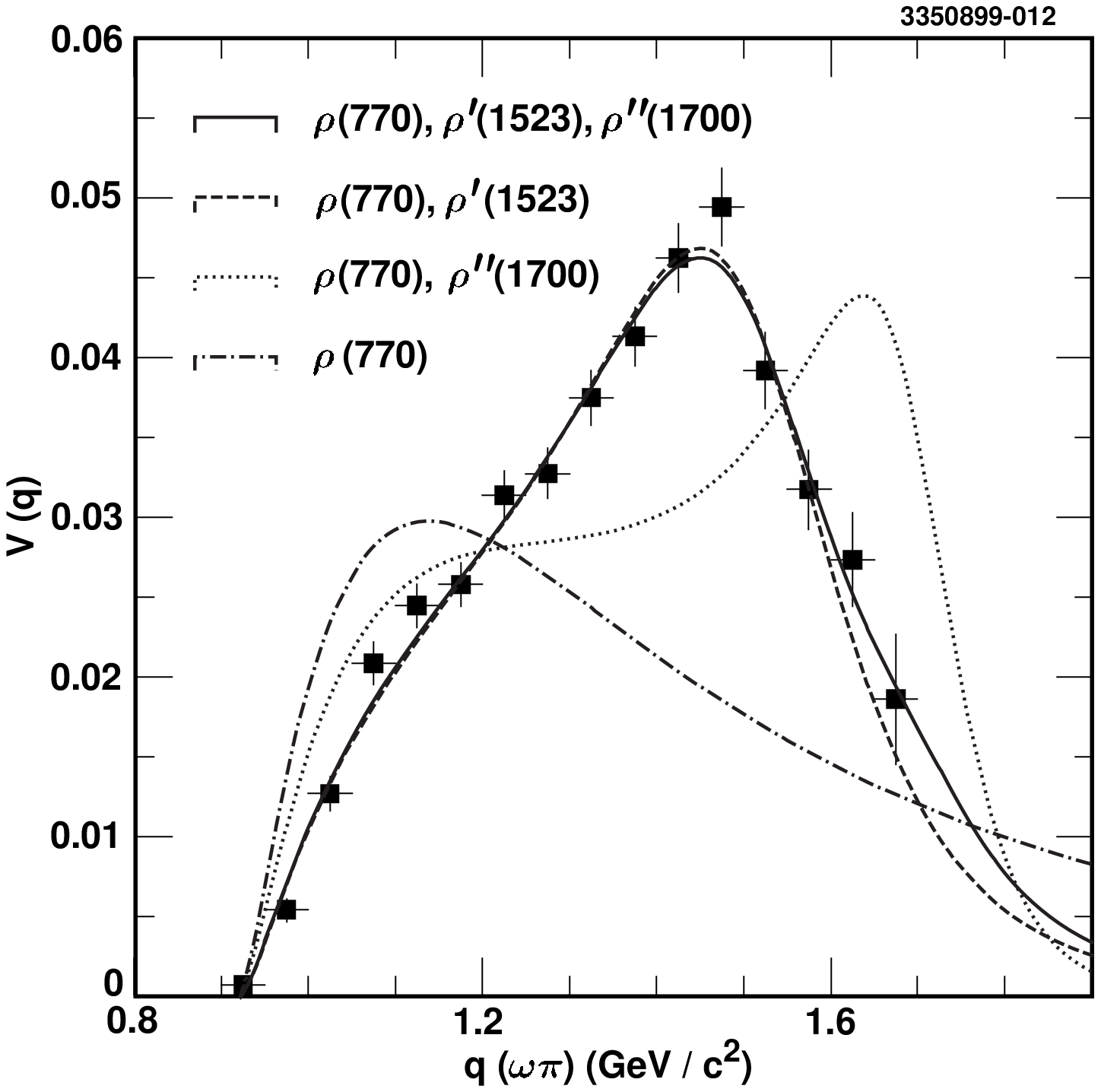,height=2.8in}
\psfig{figure=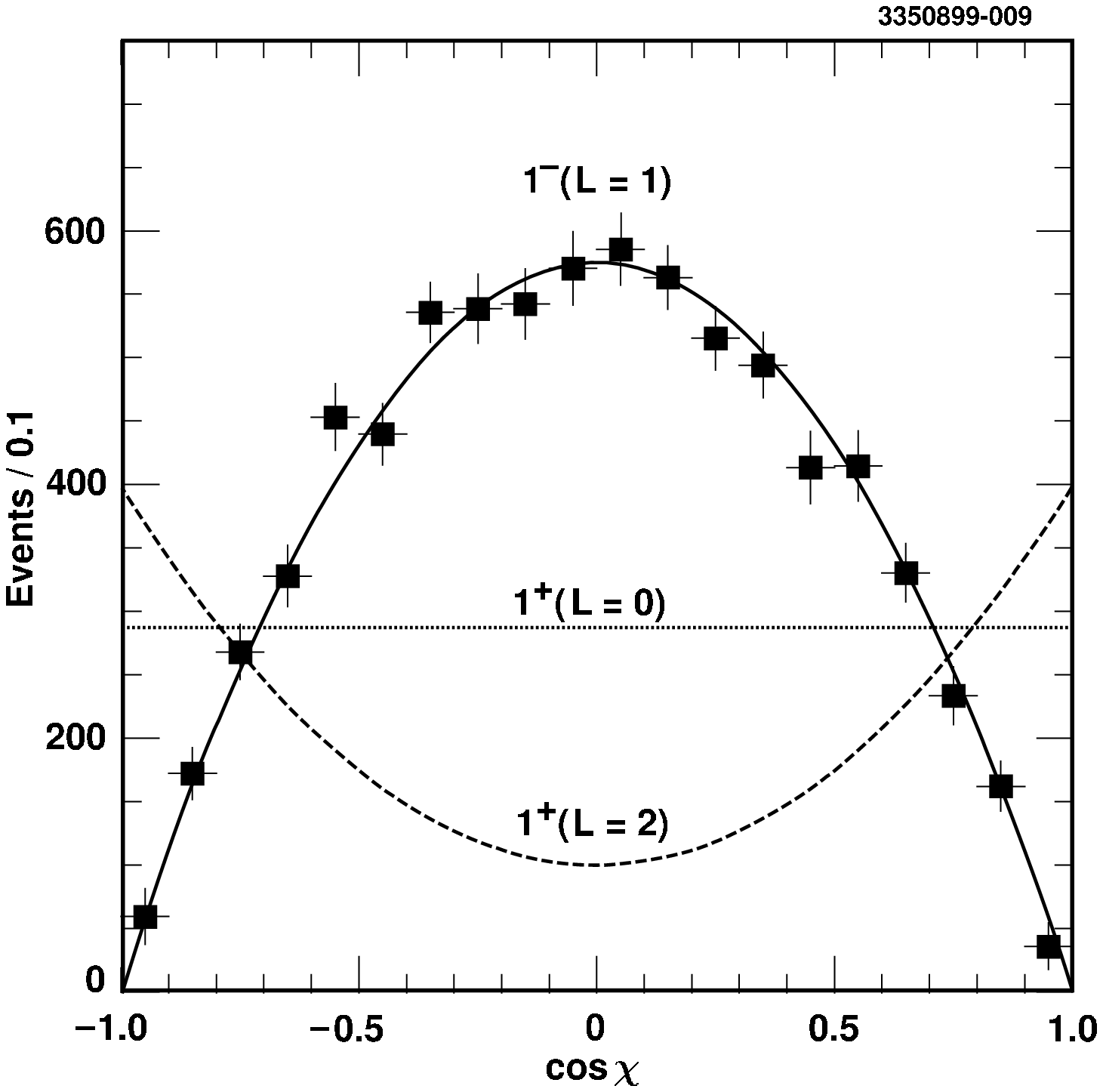,height=2.8in}
\caption{\label{fig:4pi}
Distributions from $\tau^-\to \pi^-\pi^+\pi^-\pi^0\nu_\tau$ \protect\cite{ref:CLEO4pi}.
Top: The spectral function 
$v(M(\omega\pi))$, with fits to various combinations
of $\rho$, $\rho'$, $\rho''$... resonances.
Bottom: Angular distribution sensitive to the polarization
of the $\omega$ in $X\to \omega\pi$ (points),
compared to predictions for different partial waves.
Second-class currents would reveal themselves as $L=0,2$ partial waves.
}
\end{center}
\end{figure}

\section{ Tau mass and lifetime, tau neutrino mass }

CLEO pioneered the use of kinematical constraints 
to measure the tau mass,
by observing semi-hadronic tau decays on both sides
of an event, and determining for each event
a maximum kinematically-allowed tau mass
consistent with the observed hadronic energies and momenta.
The distribution of this maximum mass exhibits a sharp
drop-off near the tau mass, and from the position
of this drop-off, CLEO measured \cite{ref:taumass}
$m_\tau = (1778.2\pm 1.4)$ MeV.

At around the same time, the BES experiment measured the tau mass
much more accurately, through a threshold scan \cite{ref:BESmass}.
CLEO turned this to an advantage;
the ``maximum kinematically-allowed tau mass''
was really a measurement of a combination of the
tau mass and the $\nu_\tau$ mass:
$m_\tau^{kin} \simeq m_\tau^{BES} - m_{\nu_\tau}^2 / m_0$
where $m_0$ is a mass parameter determined through Monte Carlo
simulation.
Making use of the BES measurement,
CLEO kinematically constrained the $\nu_\tau$ mass to be
$m(\nu_\tau) <$ 60 MeV, 95\%\ CL.
Of course, what is being constrained here is the 
effective mass of the linear combination of neutrino mass
eigenstates which couple to the tau.

CLEO-II used its vertex proportional chambers
to measure the tau lifetime \cite{ref:taulife}.
Using 1-v-3 and 3-v-3 events,
and making use of vertex and beam position information,
the tau lifetime was measured to be
$\tau_\tau = 289.0\pm 2.8 \pm 4.0$ fs,
in good agreement with more precise measurements from LEP.
CLEO-II.V introduced a precision silicon vertex detector,
but so far, that device has not been used to re-measure
the tau lifetime.

\subsection{Mass of the $m(\nu_\tau)$}
\label{ss:mnutau}

Regardless of $m(\nu_\tau)$ constraints 
from $\nu$-mixing and cosmology, 
constraining it kinematically in tau decays 
remains a worthy goal.
The ALEPH limit from 1998 still stands: 
$m^{eff}(\nu_\tau) < 18.2$ MeV (95\%\ CL).
(Again, what is being constrained here is the 
effective mass of the linear combination of neutrino mass
eigenstates which couple to the tau.)

CLEO-II has published 3 limits on $m^{eff}(\nu_\tau)$,
using different decay modes and ever-larger datasets: \\
$m^{eff}(\nu_\tau) < 32.6$ MeV, 1993, $5\pi,3\pi 2\pi^0$ \cite{ref:nutau1}, \\
$m^{eff}(\nu_\tau) < 30$ MeV, 1998, $5\pi,3\pi 2\pi^0$ \cite{ref:nutau2}, \\
$m^{eff}(\nu_\tau) < 28$ MeV, 2000, $3\pi  \pi^0$ \cite{ref:nutau3}. \\
The last two limits used the 
$M_X$  versus $E_X/E_{beam}$ technique pioneered at LEP,
where $X$ is the hadronic system in $\tau\to X\nu_\tau$ decay.
All three of these limits used a much larger data sample,
than the one available to ALEPH,
and comparable or better $M_X$, $E_X$ resolution.
The data from the second analysis listed above
is shown in Fig.~\ref{fig:nutau5pi}.

If one is motivated to improve this kinematical limit
appreciably, \eg\ to approach the $\sim$ 1 MeV level,
one needs lots of statistics,
excellent, well-understood $M_X$, $E_X$ resolution,
good spectral function models,
and most especially, a good understanding of statistics and 
systematics - there are many subtleties \cite{ref:duboscq}!

\begin{figure}[!ht]
\begin{center}
\psfig{figure=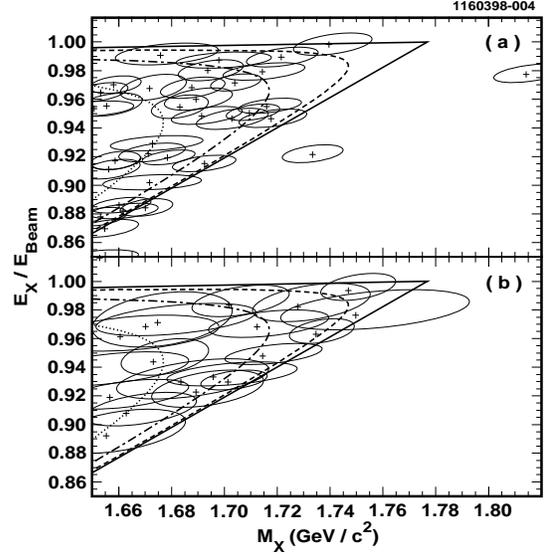,width=2.8in}
\caption{\label{fig:nutau5pi}
Distribution of invariant mass $M_X$ versus scaled energy
$E_X/E_{beam}$ for $\tau\to X\nu_\tau$,
where 
(a) $X = 5\pi^\pm$, and (b) $X = 3\pi^\pm 2\pi^0$,
in CLEO data \protect\cite{ref:nutau2} near the kinematic endpoints, 
with $1\sigma$ resolution error ellipses shown.
The solid lines show the kinematical boundary for
$m(\nu_\tau) = 0$, and the dashed contours are for
$m(\nu_\tau) = 30$, 60, and 100 MeV.
}
\end{center}
\end{figure}

\section{ CP Violation in tau decay }

A highlight of the recent work in 
tau physics from CLEO has been the search for CP violation 
in tau decay \cite{ref:ryszard}.

Of course, CP violation is not expected in
leptonic decays, in the Standard Model.
To manifest it, a process must have two
or more interfering amplitudes,
with relatively complex phases
(as in the CKM matrix in the quark sector).
In the Standard Model,
the $\mu$ and $\tau$ decay via one amplitude:   
$\tau^-\to W^-\nu_\tau$,
a process that has been well studied at CLEO and LEP.
Also, the charged leptons cannot undergo 
particle-antiparticle oscillations.

To produce CP violation in tau decay,
we can add a second amplitude,
such as a charged Higgs: $\tau^- \to H^-\nu_\tau$.
Endow it with a complex coupling $\Lambda$ with a phase 
which flips sign under CP,
and a strong phase (supplied by the $W\to \rho$ or $K^*$ 
Breit-Wigner propagator) which does not.
CLEO has recently searched for CP violation
due to such a mechanism,
in two analyses.

In the first, the decay $\tau^- \to (K\pi)^-\nu_\tau$
(or its charged conjugate) is reconstructed
on one side of the event. The other tau decay is used only
to tag the event as $\tau^+\tau^-$.
CP violation would manifest itself if the 
$K$ momentum vector lay preferentially on one side
of the plane formed by the $e^+ \tau^+$ momentum vectors;
this effect is manifestly SU(3)$_f$ violating.

In the second, both sides of the event are required
to go to $\rho\nu_\tau$:
$\tau^+\tau^-\to (\rho^-\nu_\tau)(\rho^+\overline{\nu}_\tau)$.
The $\tau\to \rho\nu_\tau$ decays are used to analyze the spin
orientation of each tau, and one looks for net transverse spin
polarization; a manifestly isospin-violating, as well as 
CP-violating effect.

In both cases, maximal information about the presence of 
CP-violating terms in the decay rate (in the context
of a model containing a scalar charged Higgs with a complex
coupling $\Lambda$) is extracted by defining an optimal
CP-violating observable, and looking for an asymmetry
in the distribution of that observable.

No such asymmetry was seen, and (model-dependent)
limits were set on $\Im(\Lambda)$: \\
$-0.172 < \Im(\Lambda) < 0.067$  at 90\%\ CL \\
in the $\pi^\pm K^0_S \nu_\tau$ -vs- any-tag analysis \cite{ref:CPKpi},
and \\
$-0.033 < \Im(\Lambda) < 0.089$  at 90\%\ CL \\
in the $\rho^+\bar{\nu}$ -vs- $\rho^-\nu_\tau$ analysis \cite{ref:CPrhorho}.


\subsection{Anomalous $\tau^+\tau^-$ production: dipole moments }
\label{ss:dipole}

As mentioned in section \ref{ss:production},
it is of interest to search for anomalous $\tau^+\tau^-$ production
at energies much below the $Z^0$, parameterized by
anomalous dipole moments.
The best sensitivity to anomalous dipole moments
can be obtained by studying the spin correlations
in $e^+e^- \to \gamma^* \to \tau^+\tau^-$ events.

Searches have recently been made by ARGUS \cite{ref:ARGUSEDM}
and Belle \cite{ref:BELLEEDM}.
Sadly, CLEO has not published on this subject.

If the tau dipole moments are not anomalously large,
then (far below the $Z^0$ peak) the taus in 
$\tau^+\tau^-$ events have very small net spin polarization.
However, their spin polarizations are almost 100\%\ correlated,
in all three dimensions.
This is interesting to measure, if only as as a test of QED.


The nature of spin correlations is a strong function of beam energy.
At 10 GeV, tau pairs are produced via $e^+e^-\to \gamma^*$.
This is parity-conserving, 
and at at energies where the taus are relativistic
but not extremely so. Both longitudinal and transverse 
spin correlations are maintained,
but there is no net spin polarization.

At LEP-I, tau pairs are produced via $e^+e^-\to Z^0$.
This is parity-violating,  
and at energies where the taus are extremely relativistic.
Both longitudinal and  transverse
spin correlations are maintained, 
but the large boost makes it nearly impossible to measure the
transverse spin polarizations;
and there is a net longitudinal spin polarization
which is well established at LEP.

Near $\tau^+\tau^-$ threshold, tau pairs are produced via $e^+e^-\to \gamma^*$.
Again, this is parity-conserving, but the taus are nearly at rest.
The spin correlations are dominantly along
the direction of the $e^+e^-$ beam axis.
Again, there is no net spin polarization.

These differences lead to significant and interesting
differences in the way anomalous dipole moments and CP violation
are manifested in tau pair production,
and different optimal observables must be defined
in order to best observe it.

\begin{figure}[!ht]
\begin{center}
\psfig{figure=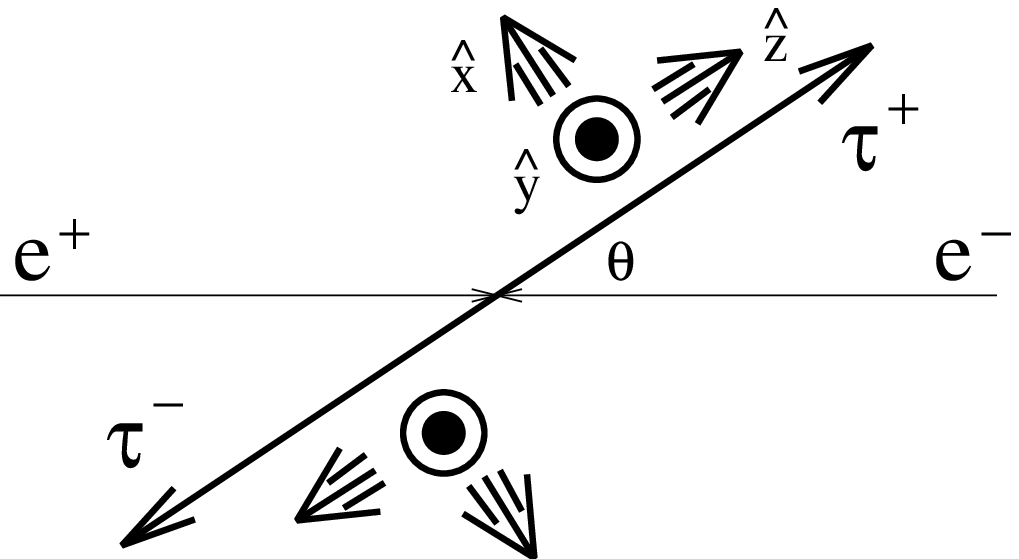,width=2.8in}
\psfig{figure=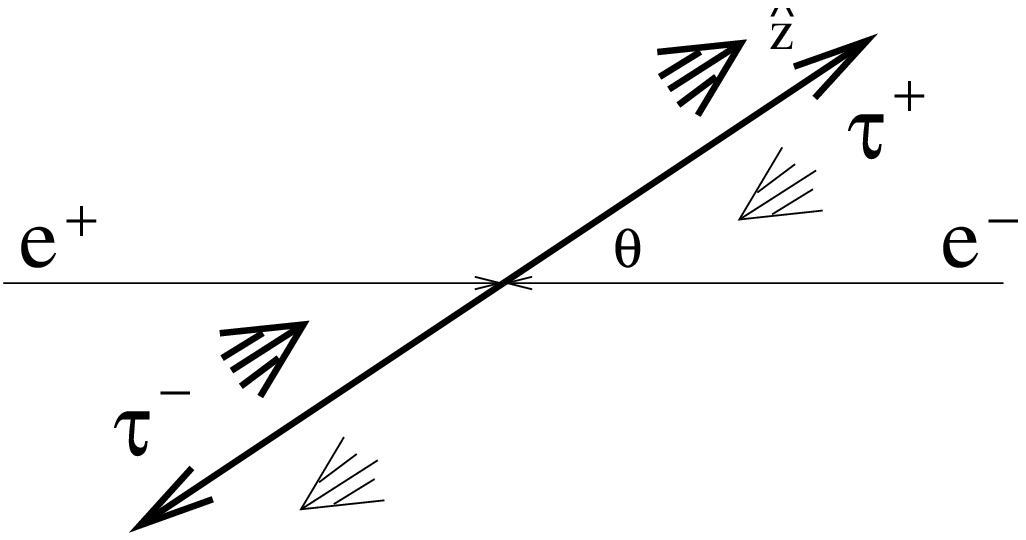,width=2.8in}
\psfig{figure=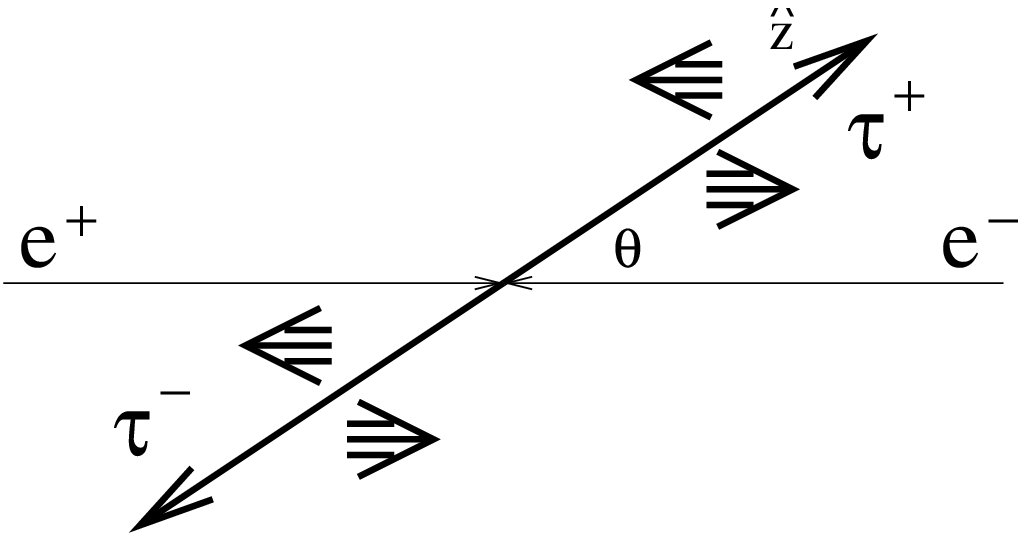,width=2.8in}
\caption{\label{fig:spincors}
Illustration of tau spin correlations in $e^+e^-$ collisions
at 10 GeV (top), near the peak of the $Z^0$ (middle),
and near threshold (bottom).
}
\end{center}
\end{figure}

\section{ Tau physics at CLEO-III and at CLEO-c}

The study of tau physics continues in the CLEO-III
era, where our RICH detector permits a more precise
study of modes containing kaons \cite{ref:feng}.

CLEO-III has $9 \times 10^6$ produced $\tau^+\tau^-$ 
near 10 GeV,  with good $K/\pi$ separation.
There is also $\sim 3\times 10^6$ fb$^{-1}$ 
collected on or near the peaks of the on $\Upsilon(nS)$
resonances ($n = 1, 2, 3$) 
which will yield measurements of $\BR(\Upsilon(nS) \to \tau^+\tau^-)$.
Other topics in tau physics which will be pursued include:
\bi
\ib rare decays, modes with kaons, precision measurements;
\ib More tests of CP in $\tau$ system:
    $h_{\nu_\tau} = - h_{\bar{\nu}_\tau}$;
\ib Rare decays may  be seen: \eg, 2nd class currents;
\ib Limits (observation?) on LFV decays;
\ib anomalous (\eg, CPv) couplings in weak decay or in QED production;
\ib exotica (\eg, $\tau^-\to\pi^-\nu_{heavy}$, $e^- G^0$);
\ib continued testing and development of models of meson dynamics 
    as a guide towards more fundamental theory: 
    structure of $\tau\to 4\pi\nu_\tau$, $K3\pi\nu_\tau$, 
              $\eta2\pi\nu_\tau$, $\eta3\pi\nu_\tau$, \etc.
\ei 

Within the next year, CESR will make the transition
to CESR-c, operating in the $E_{cm} \sim 3-5$ GeV region.
The suitably modified CLEO-c experiment will collect data
on the $\psi(nS)$ resonances, and, hopefully, 
near/at $\tau^+\tau^-$ threshold.
All the physics topics on the CLEO-III list above
will also be accessible to CLEO-c,
but with unique kinematical constraints \cite{ref:cleoc}.

When the taus are produced near threshold,
decays like $\tau\to\pi\nu_\tau$ will produce a 
monochromatic pion, which will tag $\tau^+\tau^-$ events
with good efficiency and virtually no background.
It should be possible to measure branching fractions
with sub-1\%\ precision.
It should be possible to obtain greater precision
in measurements of the Michel parameters, especially
the low-energy parameter $\eta$.
The unique spin correlations near $\tau^+\tau^-$ threshold
will permit new tests of QED, and facilitate searches
for anomalous couplings.
Threshold scans could result in measurements of 
the tau mass to a precision of 0.1 MeV.
It may also be possible to limit $m(\nu_\tau)$ kinematically
at the $\sim$ 10 MeV level.

The CLEO-c tau-charm factory has a bright future in tau physics!

\section{Summary}

The CLEO Collaboration has published numerous studies of 
the physics of the tau lepton and its neutrino,
and looks forward to continuing this work with 
data from the CLEO-III detector, and into the CLEO-c era.

The author would very much like to thank the 
organizers of Tau02 for their hospitality,
but unfortunately, he was not able to avail himself
of it, because personal problems prevented him 
from attending in person. He is grateful for the opportunity
to give his presentation remotely.


\end{document}